# On the experimental investigation of the electric and magnetic response of a single nano-structure

P. Banzer,[1,2,3,*] U. Peschel,[2,3] S. Quabis,[2] and G. Leuchs[1,2,3]

[1] *Max Planck Institute for the Science of Light, Guenther-Scharowsky-Str. 1, D-91058 Erlangen, Germany*

[2] *Institute of Optics, Information and Photonics, University Erlangen-Nuremberg, Staudtstr. 7/B2, D-91058 Erlangen, Germany*

[3] *Cluster of Excellence, Engineering of Advanced Materials, University Erlangen-Nuremberg, Nägelsbachstr. 49b, D-91052 Erlangen, Germany*

*\*peter.banzer@mpl.mpg.de*

**Abstract:** We demonstrate an experimental method to separately test the optical response of a single sub-wavelength nano-structure to tailored electric and magnetic field distributions in the optical domain. For this purpose a highly focused y-polarized $TEM_{10}$-mode is used which exhibits spatially separated longitudinal magnetic and transverse electric field patterns. By displacing a single sub-wavelength nano-structure, namely a single split-ring resonator (SRR), in the focal plane, different coupling scenarios can be achieved. It is shown experimentally that the single split-ring resonator tested here responds dominantly as an electric dipole. A much smaller but yet statistically significant magnetic dipole contribution is also measured by investigating the interaction of a single SRR with a magnetic field component perpendicular to the SRR plane (which is equivalent to the curl of the electric field) as well as by analyzing the intensity and polarization distribution of the scattered light with high spatial resolution. The developed experimental setup as well as the measurement techniques presented in this paper are a versatile tool to investigate the optical properties of single sub-wavelength nano-structures.

---

## 1. Introduction

Since many years, polarization tailored beams such as azimuthally or radially polarized beams play an important role in many fields of applications. Especially, when strongly focused they show non-trivial field distributions in the focal region exhibiting spatially separated longitudinal and transversal electric and magnetic field components [1, 2, 3, 4, 5] depending on the initial polarization state. Therefore, possible application areas range from optical confocal microscopy [6], optical trapping and tweezing [7] and material processing [8] to more fundamental research for instance probing single molecules and their orientation [9, 10]. Additionally, related orbital angular momentum modes are even used in quantum optics to investigate hyperentanglement and squeezing [11, 12].

All these fields of research have one key feature in common. They take advantage of the extraordinary spatial polarization structure as well as the intensity distribution of such modes. More recently, the extraordinary properties of azimuthally and radially polarized light were also used to investigate metallic nano-structures, thus introducing polarization tailored modes to the rising field of plasmonics. For instance, azimuthally and radially polarized light beams were used to study the transmission properties of sub-wavelength circular apertures in thin metal films [13]. It was also shown just recently, that polarization tailored light can be used very successfully to characterize the orientation and shape of metallic sub-wavelength particles by measuring the back-scattered light [14, 15]. Especially, for sensing the shape and orientation of nano-particles, cylindrically symmetric polarization states such as radial and azimuthal polarization modes are excellent tools, because the single particles can interact differently with the electric field components depending on their shape and position relative to the beam. The interaction between a metallic nano-particle with the overlapping field distribution is governed by the excitation of electric and magnetic multipoles, leading to resonant scattering depending on the wavelength and the particle geometry. In general, also the incident optical field can be decomposed into various electric and magnetic multipoles. The light emitted by the nano-structure in turn consists of different vector spherical waves. The relation between the incident and emitted multipoles can be described by a scattering matrix, which characterizes the optical response of the nano-structure. The electric dipole response constitutes just one, but for nano-structures in the optical domain usually dominant entry. The term describing the strength of a magnetic dipole created by an incident magnetic dipole wave is usually much smaller and for atoms often comparable with the electric quadrupole contributions. Therefore, order and strengths of the excited multipoles depend strongly on the field pattern which is overlapping with the investigated nano-structure. However, by shaping the nano-structure in a certain way, cross-coupling between the multipole contributions of the overlapping excitation field and the excited multipoles in the nano-structure can be realized. For instance, a homogeneous linear electric field can induce a non-vanishing magnetic dipole moment, if the structure is properly shaped. Depending on the shape of the investigated nano-structure, it can thus also be very advantageous to adapt the structure of the exciting field to the possible interaction scenarios as well as to the achievable and detectable parameters in the setup. For instance, if a magnetic dipole excited in a nano-structure shows a large overlap with an azimuthal polarization distribution in emission. To detect its emission in the far-field along the axis of oscillation, the choice of an appropriate excitation field distribution is favorable. It has to be easily and uniquely distinguishable from the field distribution re-emitted by the excited nano-structure.

In this paper we report on the first demonstration of a technique for investigating the nature of single SRR resonances by means of polarization tailored light beams. Due to the complexity of the scattering problem the identification of an individual non-dominant coefficient of the scattering matrix in terms of incident and emitted multipoles is not an easy task. A complete characterization requires exciting the nano-structure with different field distributions and analyzing the local strength of interaction as well as the spatial structure of the light emitted by the nano-structure after being excited. For that purpose, we expose the nano-structure to different field distributions and measure the coupling strength. Additionally, we analyze the vectorial properties of the reflected light with high angular resolution. Furthermore, we adapt the excitation field to the intrinsic properties of the structure under investigation.

For our investigations we have chosen a single sub-wavelength split-ring resonator (SRR) as a model system, which is well known to posses a weak but non-vanishing magnetic dipole contribution, if excited properly [16, 17, 18, 19, 20]. SRRs interact strongly with the electric field if driven at resonance and exhibit a strong electric dipole contribution as well. For the investigation of the optical properties of a single SRR we have adapted the incident polarization mode. We use a polarization mode with Hermite-Gaussian intensity distribution, such as a y-polarized $TEM_{10}$-mode, which does not have cylindrical symmetry. When focused strongly a y-polarized $TEM_{10}$-mode shows a non-trivial field distribution in the focal plane exhibiting spatially separated longitudinal magnetic as well as transverse electric and magnetic field components. This allows for selective excitation of specific resonance types of the single SRR depending on the position and orientation of the SRR relative to the beam in the focal plane.

## 2. Investigated model system: a single sub-wavelength split-ring resonator

To demonstrate our experimental method we choose a single split-ring resonator (SRR) with sub-wavelength dimensions (see Fig. 2(b)), because those structures are supposed to have pronounced resonances of different types as well as a significant and resonant magnetic response in the optical domain depending on there orientation relative to the excitation field. Note that here the term magnetic serves to classify the scattering response but gives no hint on the internal dynamics inside the nano-structure. The currents inside the SRR are still driven by an electric field. Rockstuhl et al. have therefore proposed to classify these resonances on the basis of plasmonic modes [21]. Following this notation we focus exemplarily on the lowest order resonance, where just half the wavelength of a plasmon fits around the SRR. Hence, the induced currents should lead to eddy currents resulting in a maximum magnetic moment (see inset in Fig. 3(a)). Therefore, this resonance is sometimes called magnetic. However, such a phrase has to be used with care. For electro-magnetic fields a differentiation in terms of electric and magnetic response has only limited meaning because e.g. the temporal derivative of the magnetic field is equivalent to the curl of the electric field according to Maxwell's equations. The charges inside the nano-structure are always driven by the electric field. Furthermore the magnetic moment of the induced current distribution seems to weaken in the optical domain. On the basis of numerical simulations Soukoulis et al. [16] have predicted, that the resonances of SRRs have a diminishing magnetic dipole contribution, if the SRR deviates from the ideal geometry of the closed loop structure used at microwave frequencies. In particular, a reduction of the two SRR arms which is used to shift the resonance to the optical domain is predicted to reduce the magnetic response or equivalently the $(\nabla \times \mathbf{E})$ response.

The investigated SRR (see Fig. 2(b)) is fabricated in-house using standard focused-ion-beam techniques. Well separated and individually addressable SRRs are patterned into a 30 nm thin gold film on a 170 $\mu$m glass slab. The dimensions of the investigated single SRR are also shown in Fig. 2(b).

## 3. Tailored polarization pattern for investigating single sub-wavelength nano-structures

For investigating the optical properties of single nano-structures, we have chosen a y-polarized $TEM_{10}$-mode for the reasons discussed above. Sketch a) of figure 1 illustrates the intensity pattern of a collimated y-polarized $TEM_{10}$-mode at a wavelength of 1.525 $\mu$m used in the experiments. The electric field components point in the y-direction with a Hermite-Gaussian intensity distribution, where the magnetic field points in the x-direction. Between the lobes, a polarization singularity is formed due to a phase jump of pi. When strongly focused (see figure 1(b)), the magnetic field vectors are tilted toward the optical axis (z-axis). In the focus they add up pair wise to a strong longitudinal magnetic field component (z-direction). The zero of the electric field on the optical axis persists (see figure 1(c) and (d)). Furthermore, transverse electric and magnetic field components still exist off-axis [2]. A non-homogeneous polarization distribution is achieved in the focal plane. Figure 1(c) and (d) show the cross section through the electric and magnetic energy densities along the x- and y-axis in the focal plane of a microscope objective with high numerical aperture. A pure longitudinal magnetic field is present on-axis. The distributions of the electric and magnetic field components are calculated using the Debye method.

An important reason for the choice of polarization tailored light originates from the versatility of this technique as well as the resonance type of the investigated single SRR. The resonance observed experimentally around 1.525 $\mu$m (see Fig. 3(a)) is normally interpreted as a magnetic resonance with a magnetic dipole oscillating perpendicular to the SRR plane. To investigate for instance this resonance type in detail, we have to test the response of the single SRR to an overlapping field distribution containing no electric dipole contribution thus not stimulating the dominant electric dipole response of the SRR. We use a highly focused y-polarized $TEM_{10}$-mode to realize the intended coupling. As already mentioned, this polarization state leads to a strong z-component of the magnetic field on-axis, which is directly connected to a curl of the electric field in the focal plane. In the experiment, the longitudinal magnetic field component points perpendicular to the structure plane for normal incidence. Thus, the equivalent electric field distribution lies in-plane and should induce eddy currents in the resonantly shaped SRR depending on the leading to the excitation of a magnetic dipole. Due to the non-cylindrical symmetric geometry of the incoming beam, also in the focal plane the electric and magnetic field distributions do not exhibit cylindrical symmetry (see Fig. 1). Therefore, depending on the orientation of the SRR relative to the beam (and the excitation wavelength), the coupling strength will change. Additionally, the sub-wavelength dimensions of the investigates nano-structure and the tailored field distribution in the focal plane allow also for the realization of various coupling scenarios. For instance, the single SRR can be placed in one of the lobes of the beam thus being exposed to a strong and almost homogeneous electric field. To enhance the sensitivity of the measurements, a y-polarized $TEM_{10}$-mode was chosen in the presented experiments instead of a azimuthally polarized beam. The reasons for that choice will be discussed in detail later on.

## 4. Experimental setup and measurement technique

To investigate the optical properties of single nano-structures we use a special setup (see Fig. 2(a)). For performing spectrally resolved measurements a ultra-broadband and tunable light source (LS; pumped highly nonlinear photonic-crystal fiber combined with a liquid-crystal spectral filter (LCSF); spectral range: 0.8 $\mu$m - 1.8 $\mu$m; filter band-width: 5 - 8 nm) is used. To investigate the resonance type and response of single SRRs, a linearly polarized Gaussian beam as well as polarization tailored light (e.g. a y-polarized $TEM_{10}$-mode; see figure 1(a)) is used. For measurements with polarization tailored light we firstly generate an azimuthally polarized doughnut mode by an electronically controlled polarization conversion device (LCPC), namely a liquid-crystal cell (for basic principle see [22]). Because a pure azimuthally polarized ring-mode is a superposition of a x-polarized $TEM_{01}$ and a y-polarized $TEM_{10}$-mode. We can achieve the addressed polarization state easily in the setup by introducing a linear polarizer behind the liquid-crystal device. Due to the working principle of the liquid-crystal device and its intrinsic imperfections the generated mode is not pure. Thus, also higher-order modes are present which also distort the shape of the main lobes. Nevertheless, the diameter of the beam is adjusted in that way, that the additional maxima of the higher-order modes and impurities are cut by the entrance pupil of the microscope objective. Note, that this original shape of the incoming field at the entrance pupil of the microscope objective is taken into account when simulating the expected field distribution and field components in the focus (see Fig. 1). The LCPC is removed when measurements with a linearly polarized Gaussian beam are performed.

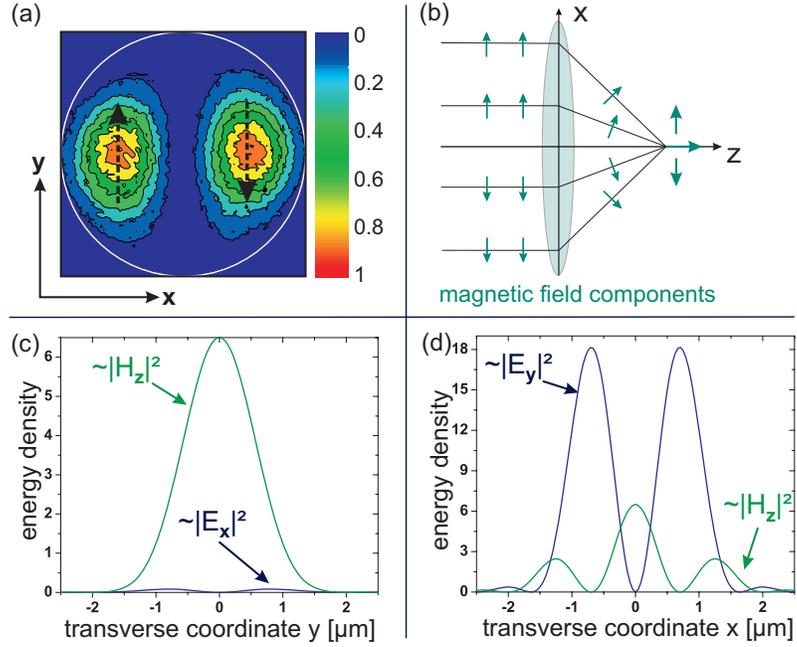

Fig. 1. (a) Experimentally recorded intensity distribution at a wavelength of 1.525 $\mu$m present at the entrance pupil of the focusing microscope objective in the setup. The white circle represents the diameter of the entrance pupil of the microscope objective (3.6 mm) in comparison to the intensity distribution (drawn to scale). The black dashed arrows show a snap-shot of the polarization. (b) Focusing scheme (x-z-plane) shown for the magnetic field vectors of a y-polarized $TEM_{10}$-mode propagating in z-direction. A strong longitudinal magnetic field is formed on-axis in the focal plane, whereas the electric field is zero. Calculated energy density of the longitudinal magnetic field $H_z$ in the focal plane. The small tilt of the field distribution is caused by the slight asymmetry of the incoming field. Cross-sections showing the calculated energy densities along (c) the nodal line (y-direction) and (d) perpendicular to the nodal line (x-direction) through the focal plane of a y-polarized $TEM_{10}$-mode; $\frac{\mu_0}{2}|H_z|^2$ (solid green line), $\frac{\varepsilon_0}{2}|E_x|^2$ (solid blue line in (e)), $\frac{\varepsilon_0}{2}|E_y|^2$ (solid blue line in (f)) shown for a NA of 0.9 and a wavelength of 1.525 $\mu$m. The same arbitrary energy density unit is used in all graphs so that $\frac{\mu_0}{2}|H_z|^2$ and $\frac{\varepsilon_0}{2}|E_x|^2$ as well as $\frac{\varepsilon_0}{2}|E_y|^2$ can be compared. Note the different scales in (c) and (d). The generated transverse magnetic field components are not shown here.

The generated mode is then guided by four mirrors (M; all aligned under 45 degrees relative to the impinging beam top down through a additional linear polarizer into a microscope objective (MO) with high numerical aperture (NA = 0.9) and strongly focused on the single nano-structure (split-ring resonator: SRR) siting on a glass substrate (GS). The microscope objective also collects the reflected and back-scattered light. A non-polarizing beam-splitter (NPBS) guides the light to a photodiode (PD) which measures the total intensity of the back-scattered and reflected light. Likewise the angular distribution of intensity and polarization of the reflected light is monitored with an IR camera (CAM). The sample is mounted onto a 3D piezo table. Below the sample an immersion microscope objective (NA = 1.3) is matched with immersion oil to the glass substrate, collecting the transmitted as well as forward scattered light. The collimated beam behind the second microscope objective is then impinging onto a photodiode (PD) to measure also the total intensity of the forward-scattered and transmitted light. An additional camera can be used to record also the spatial distribution of the transmitted light (not shown here).

By displacing the single SRR in the focal plane (by means of the 3D piezo-table with sub-nanometer resolution), different coupling scenarios can be realized. For every position of the single nano-structure in the focal plane relative to the beam the total intensity of the forward-scattered and transmitted light and the back-scattered and reflected light is measured simultaneously by means of the two photodiodes. Depending on the position of the single SRR in the focal plane relative to the optical axis different coupling scenarios can thus be achieved. On the one hand, the coupling with a transverse electric field which is almost homogeneous across the SRR plane. On the other hand, the coupling with a electric field distribution having a phase jump of pi across the SRR plane, which always linked to a longitudinal magnetic field component, can be realized. Additionally, the strength of the coupling and interaction strongly depends on the orientation of the SRR with respect to the local field as well as on the wavelength of excitation.

Using the experimental setup explained above gives us full access to the light emitted by the nano-structure. We can record its far-field, because the light which is forward- and back-scattered is collected by the two microscope objectives. This permits for the analysis of the polarization as well as of the spectral properties of the scattered light. For the given geometry of the single SRR under test the primarily investigated resonance is located around 1.525 $\mu$m (see Fig. 3). Therefore, the setup is optimized for this wavelength.

## 5. Experimental results

First, we measure the spectral response of a single SRR with a linearly polarized and highly focused Gaussian beam to specify the position of respective resonances of the SRR. If a linearly polarized Gaussian beam is focused strongly, also longitudinal as well as crossed in-plane electric field components are created [23]. All these field components lie off-axis and are very weak for the chosen numerical aperture of the focusing microscope objective lens. Thus, if the SRR is placed centrally in the focal spot (on-axis) its excitation is dominated by the transverse electric field. In contrast, the crossed in-plane and longitudinal electric field components have to be taken into account when analyzing the light emitted by the nano-structure. Furthermore, it is worth to mention, that the diameter of the Gaussian beam at the entrance aperture of the microscope objective for different wavelengths is changed in that way, that the diameter of the beam in the focus is kept almost constant.

For each wavelength we scan a defined area of the sample including one SRR and register the minimum value of transmission and the maximum value of reflection achieved for the SRR being centered in the focus. The SRR lies in the plane perpendicular to the **k**-vector of the incoming field. The acquired measurement data points in transmission and reflection are normalized to the transmission through the glass substrate and the reflection at the air-glass interface, respectively. Additionally, several consecutive measurements were averaged to improve the effective signal-to-noise-ratio. Figure 3 shows the experimental results for transmission and reflection for the electric field of the incoming Gaussian beam being parallel (Fig. 3(a)) or perpendicularly (Fig. 3(b)) polarized to the SRR arms. Consistent with previous experiments on SRR arrays also the optical response of a single SRR is dominated by two resonances (in the investigated wavelength range) and depends strongly on the polarization of the incident light [17, 18, 19, 24, 25, 26].

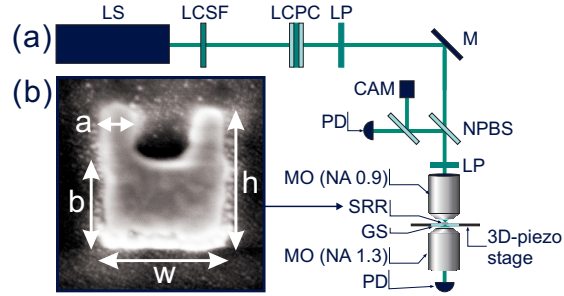

Fig. 2. (a) Experimental setup for investigating the response of a single nano-structure (SRR). LS: broadband light-source, LCSF: liquid-crystal spectral filter, LCPC: liquid-crystal polarization converter, LP: linear polarizer, M: mirrors, NPBS: non-polarizing beam splitter, MO: microscope objective (focusing MO) with an NA of 0.9; oil-immersion MO with an NA of 1.3 for collecting the transmitted light), PD: photodiode, CAM: camera, SRR: single split-ring resonator, GS: glass substrate. (b) Electron-micrograph of the investigated SRR design patterned into a 30 nm thick gold film (on a 170 $\mu$m thick glass substrate) using standard focused-ion-beam technique. SRR dimensions: h = 250 nm, w = 220 nm, b = 155 nm, a = 55 nm.

If the linear polarization is chosen such that the electric field is pointing perpendicular to the SRR arms a clear dip in transmission together with an increased reflection is observed around 1.525 $\mu$m (see Fig. 3(a)). At this wavelength, the lowest order resonance is excited for the given geometry of the tested SRR, which was interpreted as magnetic resonance in [17], [18], and [24]. Note that the excitation is solely due to the electric field, which causes the current to flow along the lower bar of the SRR. The effect of the magnetic field is negligible in this case because its longitudinal component vanishes in the center of a linearly polarized beam. By rotating the linear polarization of the exciting light (parallel to the SRR arms), the coupling scenario is changed. As already known from the literature [17, 18, 24], the so-called magnetic resonance vanishes if the exciting electric field is parallel to the arms of the SRR. Another resonance appears instead around 1.2 $\mu$m (see Fig. 3(b)). This second order resonance has a completely different current distribution with a zero at the bar center explaining its lack of excitation for electric fields pointing along the bar.

Note that here the term magnetic resonance serves to classify the scattering response but gives no hint on the possible types of excitation. In both cases the currents inside the SRR are still driven by an electric field. In the following, we focus on the investigation of the resonance observed around 1.525 $\mu$m (see Fig. 3(a)).

To further characterize the properties of the excited resonance, we now have to analyze also the spatial vectorial properties of the light re-emitted by the nano-structure for the given kind of excitation. To effectively excite the resonance, we choose the electric field of the incoming beam perpendicular to the SRR arms (see Fig. 3(a)) and read out the camera signal instead of the signal of the single photodiodes in reflection as well as transmission (see Fig. 2(a)). The light which is back-scattered by the nano-structure and reflected at the air-glass interface is collected and collimated by the upper microscope objective. The intensity pattern of the beam cross-section is now recorded with the camera (without using any imaging lens in front of the camera). By placing a linear polarizer in front of the camera we are able to analyze the polarization distribution in the beam. The open design of the setup also permits using the same approach for the transmitted and forward-scattered light. In the following, only reflection measurements are presented. Using the scheme described above we firstly record and analyze the camera image for the case, when the linearly polarized Gaussian beam is just focused onto the glass substrate not overlapping with the single SRR. This gives us a reference of the background being present, when the beam overlaps with the SRR. The observed beam cross-section has again a Gaussian shape as expected, if no additional polarizer is used in front of the camera. The same holds for a polarizer being parallel to the polarization of the incoming field (y-polarized). The situation changes completely, if the polarizer is rotated by 90 degrees (crossed polarizer; x-polarized light is passing through the polarizer). The corresponding camera images are shown in Fig. 4(a) and (c). A four-lobe pattern is detected. The electric field components of this pattern are oriented perpendicular to the incoming polarization. Thus, the highly focused linearly polarized Gaussian beam contains crossed electric field components after being reflected at a plane glass surface and collimated by the microscope objective. The reason for the non-perfect reconstruction of the incident beam in terms of the observed four-lobe pattern as well as the sub-structure is the angular dependent Fresnel reflection of its plane wave components at the glass surface. In particular azimuthal polarization components, which are responsible for beam distortions in the focal plane are more efficiently reflected than their radial counterparts. Consequently, crossed polarization components finally appear in the back-reflected beam. Additionally, a part of the beam undergoes a phase jump of pi upon reflection under an angle of incidence larger than the Brewster's angle. This is believed to cause the additional fine sub-structure observed in the camera images (similar to [27]).

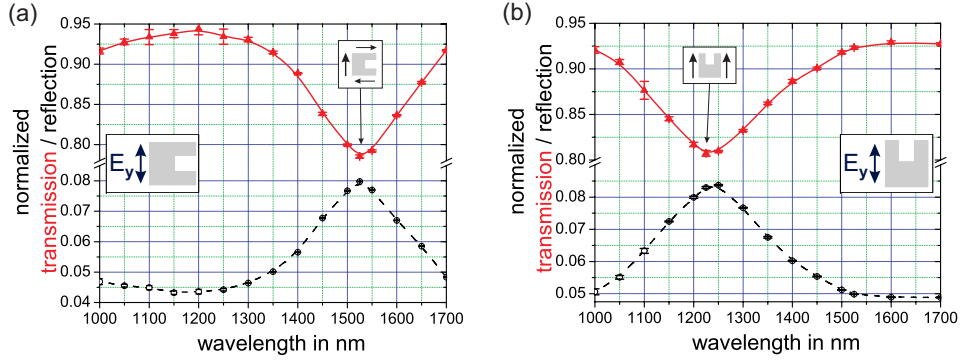

Fig. 3. Experimental resonance spectra showing the transmitted (dashed black line; black circles) and reflected (solid red line; triangles) light of a single SRR centered in a linearly polarized and highly focused Gaussian beam. Two scenarios of the polarization orientation relative to the SRR arms are shown. (a) The electric field is perpendicular to the SRR arms. (b) The electric field is oriented parallel to the SRR arms. The pronounced resonance around 1.525 $\mu$m for a polarization perpendicular to the SRR arms is the lowest order or so-called magnetic resonance. The corresponding current distributions in the SRR for the resonances observed in a) and b) are depicted by the small insets. The experimental data points for the transmission and reflection spectra are normalized to the transmission through the glass substrate and the reflection at the air-glass interface, respectively.

If the single SRR is moved into the focus of the linearly polarized beam the intensity of the observed pattern increases strongly for a parallel polarizer (not shown here). This is in line with the spectral measurements (see Fig. 3) where a maximum in reflection is observed at the given wavelength of 1.525 $\mu$m compared to the background (reflection at the glass surface). This fact already implies a strong electric dipole contribution to the investigated so-called magnetic resonance. Similarly, the pattern of the crossed polarization components (observed with a crossed polarizer in front of the camera) also changes drastically, if the SRR is moved into the focused beam. The recorded camera image in reflection is shown in Fig. 4(b). On the one hand, the intensity of the four-lobes changes [30]. Most important for the discussion here, light is detected also between the upper as well as the lower lobes (on the y-axis). This clearly indicates the presence of a weak magnetic dipole contribution to the investigated resonance. If a magnetic dipole moment is excited perpendicular to the focal plane (along the z-axis) in the nano-structure the light pattern emitted in the direction of the dipole moment observed in the far-field will have an azimuthally polarized ring shape. Due to the crossed polarizer in front of the camera the azimuthal ring shape results in a two-lobe pattern. In addition to this magnetic dipole emission, part of the light is still simply reflected at the glass substrate, because the particle is much smaller than the focal spot. Furthermore, the electric dipole mentioned above oscillating in the y-direction has also non-vanishing crossed electric field components (x-components). These field components interfere with each other forming the far-field image shown in Fig. 4(b) (see [30]). The emission pattern of an electric dipole observed through the crossed polarizer results in a 4-lobe pattern much like the pattern reflected by the substrate, resulting in zero intensity on the y-axis. Therefore, the scattered light intensity on the y-axis clearly indicates the presence of a magnetic dipole emission.

Additionally, we recorded the back-scattered light at a different wavelength (1.25 $\mu$m) to proof the observed effect being an intrinsic property of the resonance around 1.525 $\mu$m. No significant fingerprint of a magnetic dipole emission was found here (data not shown). Furthermore, the measurements were repeated for the electric field of the incoming beam being parallel to the SRR arms. In this configuration the SRR does not show a resonance around a wavelength of 1.525 $\mu$m (see Fig. 3(b)). No evidence of a magnetic dipole was found in this configuration (data not shown here). Also the electric dipole contribution is much weaker at the wavelength around 1.525 $\mu$m compared to the case of the SRR arms being perpendicular to the electric field. This is expected from the spectral measurements, where the strength of the response for both orientations of the SRR relative to the SRR arms differs strongly. For further comparison also a single rectangular nano-structure with equivalent dimensions, which does not exhibit a resonance at 1.525 $\mu$m, was tested and no significant change in the reflection pattern was observed (see Fig. 4(d)). The corresponding camera image for the reflection at the air-glass interface is shown in Fig. 4(c).

Note, that to our knowledge this is the first direct observation of the magnetic dipole contribution to a single SRR resonance.

The presented measurements already provide a detailed insight into the physics of the resonance under investigation. We further study the excitation of the resonance by using a beam with a tailored polarization pattern.

After having identified the spectral response of the nano-structure, we are now able to selectively investigate the type and character of its specific resonances. For that purpose, we use polarization tailored light, namely a y-polarized $TEM_{10}$-mode (see Fig. 1), to expose the nano-structure to a field distribution which has a large magnetic dipole contribution represented by a strong longitudinal magnetic field in the beam center. For that purpose, the single SRR is scanned through the highly focused polarization tailored beam in different ways as described below.

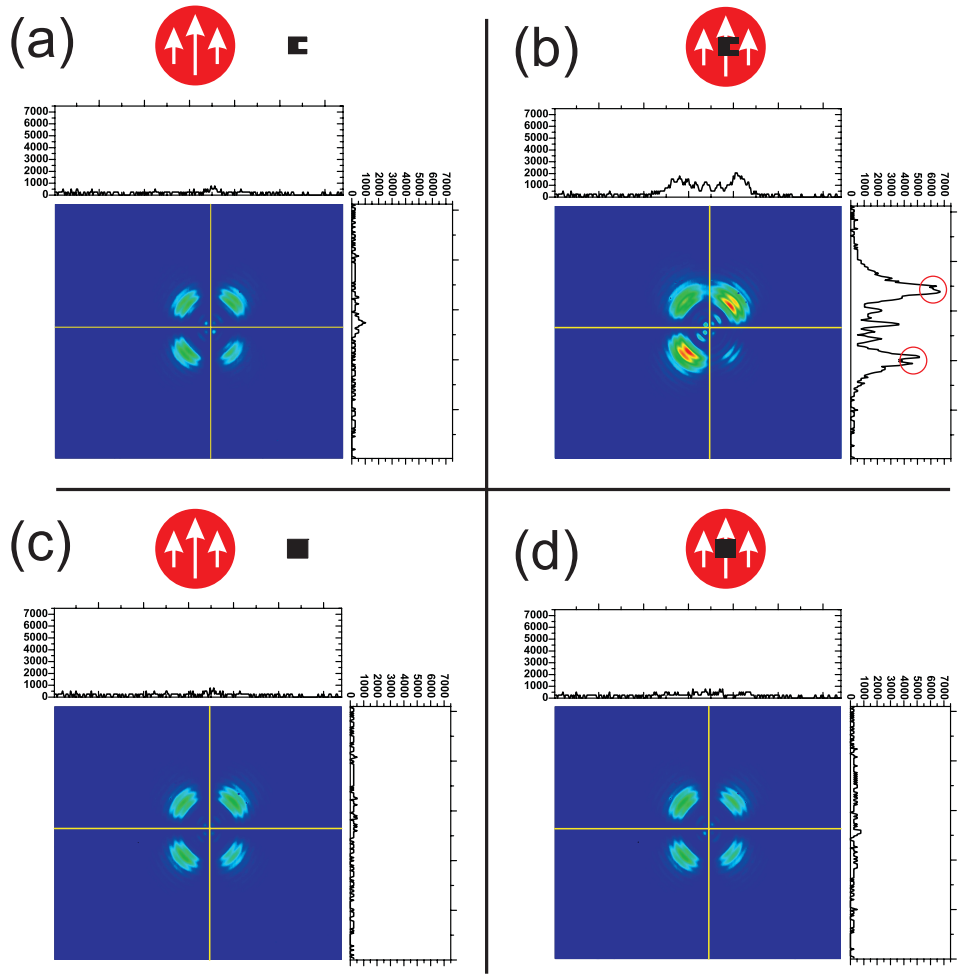

Fig. 4. Camera images of the reflected and collimated beam recorded with a bare camera sensor. The sample is illuminated with a strongly focused linearly polarized (y-direction) Gaussian beam. A crossed polarizer (x-components of the electric field can pass) is placed in front of the camera. The corresponding positions of the nano-structure relative to the beam are depicted on top of each camera image. (a) and (c): reflection at the glass substrate only; (b): SRR is placed in the beam; (d): rectangular nano-structure is placed in the beam. The cross-sections through the camera images (yellow solid lines) are shown next to the camera images. All camera images shown here were taken for a fixed wavelength of 1.525 $\mu$m. A clear fingerprint of a magnetic dipole radiation is observed in (b), if the SRR is placed in the beam (SRR arms perpendicular to the electric field).

When excited by a linearly polarized beam, the investigated resonance around 1.525 $\mu$m showed a non-vanishing magnetic dipole contribution in emission. The question arises, whether this resonance can also be excited by a strong magnetic field perpendicular to the SRR plane, which is directly connected to a curl of the electric field in-plane. For that purpose, we expose the nano-structure to a field distribution which has a large overlap with a magnetic dipole. To diversify the introduced measurement technique, we use the described polarization mode instead of azimuthally polarized light, which also exhibits a strong z-component of the magnetic field. On the one hand, we obtain an additional degree of freedom, namely the orientation of the single SRR in-plane relative to the field. Especially for the model system tested here (single SRR), which is asymmetric, this can help to gain additional information about the excitation of a nano-structure. On the other hand, the nodal line between the two main lobes of the electric field in the focus can be used as a scan line of the SRR. Due to the fact, that the expected response will be very small but the response to a linear electric field will be very high, this technique allows for averaging multiple scans, while keeping the overlap of the nano-structure with the strong linear and homogeneous electric field in the main lobes low and maximizing the overlap with the intended field distribution. Thus, the signal-to-noise ratio can be improved significantly. In addition, a y-polarized $TEM_{10}$-mode is less sensitive to distortions of the polarization pattern when reflected at the mirrors in front of the microscope objective.

In Fig. 5 different scan results are shown. The orientation of the SRR is chosen such that the SRR arms point into the x-direction, therefore being perpendicular to the y-polarized $TEM_{10}$-mode (see Fig. 5(a)). Figure 5(b) represents a color-coded two-dimensional scan result (black: lowest detected reflection; white: highest detected reflection) measured in reflection. The scan was performed using a step-size of 50 nm. The scanned field has an area of 5 $\mu$m times 5 $\mu$m. Each pixel of the scan image represents the total amount of light collected by the upper micro-scope objective and measured with the photodiode for a different position of the single SRR relative to the optical axis in the focal plane. The 2D scan result in Fig. 5(b) already shows the strong response of the single SRR to a linear electric field homogeneous over the nanostructure as expected from the measurements with a linearly polarized Gaussian beam above. This response to an linear electric field can also be quantified further by scanning the sample along a line through the main lobes in the focus (perpendicular to the nodal line; see horizontal dashed white line in Fig. 5(b)). By repeating this line scan several times while detecting the reflection signal and averaging the results, the signal-to-noise-ratio can be improved. The step-size was again chosen to be 50 nm. The length of the scan line was set to 5 $\mu$m. Note, that for all line scans presented here the relative position of the scan line (nodal line or line perpendicular to it) as well as the focus position was checked and corrected if necessary between the individual scan measurements. This procedure is required to minimize the effect of the piezo stage slowly drifting on the nanometer scale. Figure 5(c) shows the averaged results (10 line scans averaged) of scans along the line perpendicular to the nodal line and through the maxima of the main lobes. The normalization level was set to the known reflection at the air-glass interface. The insets depict the expected positions of the SRR relative to the beam for one of the maxima and the local minimum of the scan result. The response of the SRR to a linear and homogeneous electric field (SRR sitting in one of the main lobes) is by orders of magnitude stronger than for the case where the SRR is placed between the lobes on the nodal line. Although, the energy density of the z-component of the magnetic field being present on-axis is comparable in strength to the maximum electric energy density in the lobes (see Fig. 1), the difference in response is tremendous. The reflection signal between the lobes goes almost back to the normalization level. This can be explained by looking closer into the process of interaction for both coupling scenarios. For the SRR sitting at one of the maxima of the electric field lobes the linear electric field component can couple resonantly to the particle for the chosen wavelength driving the electrons along the bar (which connects the two SRR arms). Due to the resonant ring structure of the SRR, the induced currents lead to eddy currents, resulting in a maximal possible magnetic dipole as well as a strong electric dipole contributions from the electrons flowing almost linearly in the SRR bar. This is in line with the measurements of the vectorial distribution of the SRR's radiation spatially resolved using a linearly polarized Gaussian beam presented above. If the SRR is sitting on the optical axis (on the nodal line), the scenario changes completely. Although the overlap with the magnetic energy density is maximum for this constellation, induction remains weak and the induced currents stay at a very low level. Additionally, the orientation of the single SRR was chosen such that the contribution of the electric field components in the main lobes (on the x-axis) is only marginal at the given wavelength. The effect induced by the electric field components of the main lobes are canceling each other due to their opposite phase. Therefore, only a very weak coupling is expected from the overlap with these field components. Only the very weak overlap with the magnetic and the thus induced electric field is present on the y-axis (along the nodal line; see Fig. 1(c)) can lead to a resonant coupling, possibly resulting in a very small eddy current. Nevertheless, the overlap of the crossed electric field components in-plane is by 4 orders of magnitude smaller compared to the overlap of the electric field with the SRR when sitting in one of the main lobes (derived from simulation of the focal field distribution). Thus, this overlap should not lead to a significant scattering in the scan measurements.

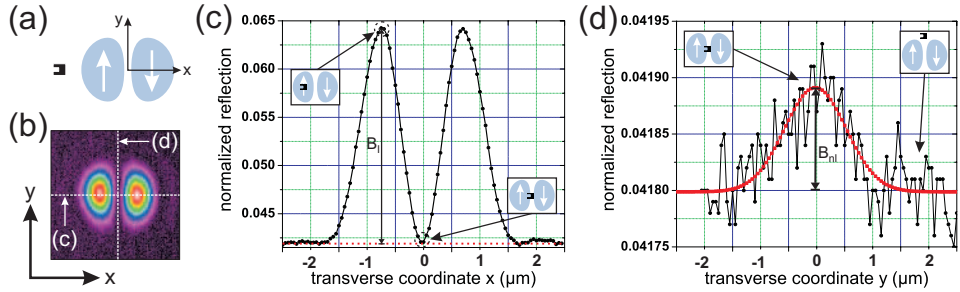

Fig. 5. Measurements of the intensities of the back-scattered and reflected light at a wavelength of 1.525 $\mu$m performed by scanning the focal plane of a y-polarized $TEM_{10}$-mode. (a) Orientation of the single SRR relative to the electric field of the incoming beam (SRR arms perpendicular to the electric field). (b) Color-coded 2D scan image showing the reflected part of the focused beam measured as a function of the relative position of the center of the focused beam and the center of the SRR (recorded in the focal plane). The scanned field has an area of 5 $\mu$m times 5 $\mu$m. The step-size was set to 50 nm. (c) Line-scan through the maxima of the main lobes of the electric field in the focus (x-direction; perpendicular to the nodal line). 10 scans were averaged and normalized to the reflection at the glass substrate. The step-size was 50 nm. (d) Averaged and normalized scan result for scans along the nodal line. 66 line scans were averaged. The red curve represents the result of the iterative nonlinear least-squares fit routine. $B_l$ and $B_{nl}$ represent the maximum response achieved for the scan perpendicular to the nodal line (SRR in one of the main lobes of the electric field) and along the nodal line, respectively.

To check if the scattering for the SRR sitting on the nodal line is vanishing or not, we perform repeated scans along the nodal line (represented by the vertical white line in 5(b)), which are averaged afterwards to reduce the noise. For this purpose, the setup was adjusted such that the scans are performed in the focal plane and along the nodal line. The nodal line is found by moving the SRR to the position of minimum reflection between the two electric field lobes of the y-polarized $TEM_{10}$-mode. This can be done with an accuracy of less than 50 nm. The reflection was scanned (duration: 3s) along the nodal line 11 times. This procedure was repeated 6 times in an attempt to reduce the effect of the piezo stage slowly drifting on the nanometer scale. Finally all 66 scans were added and normalized and the normalization level was set again to the known reflection at the glass substrate. Figure 5(d) shows the averaged scan results. Consistent with the results discussed above, the measured response of the SRR to the field distribution present along the nodal line is very weak. Therefore, to further quantify the strength of the observed coupling and proof the measured amplitude to be significant, we use a fit routine. The theoretically expected reflection function along the nodal line is

$$f(x) = A + B \times exp(\frac{-2(x-x_0)^2}{\omega^2}). \qquad (1)$$

The four parameters $A$, $B$, $x_0$ and $\omega$ were determined by fitting the function to the scan data with a nonlinear least-squares fit routine [28]. The fit curve is shown in Fig. 5(d) (red curve). The least-squares fit routine allows also for a determination of the statistical error of the fit parameters. For the scan in Fig. 5(d) we obtain $B_{nl} = 0.000092(52)$. Therefore, the amplitude for the given coupling scenario is significantly non-zero.

As a proof of principle the single SRR was rotated by 180 degrees in the x-y-plane (SRR arms point in positive x-direction) and the measurements were repeated. As a result the same values for the response discussed above were achieved and hence the symmetry of the field distribution has been be proved.

Next, the orientation of the single SRR relative to the polarization of the incoming field is changed (SRR arms point along the y-axis; see Fig. 6(a)). The wavelength is still fixed to the so-called magnetic resonance around 1.525 $\mu$m. Figure 6(b) - (d) show the 2D- and line-scan results for the given configuration. Compared with Fig. 5 we notice completely different magnitudes of reflected power. On the one hand, the response of the single SRR positioned in one of the main lobes seems to be weaker than for the rotated configuration. On the other hand, the maximal response of the particle sitting between the lobes is apparently stronger than before and furthermore not centered. To verify and quantify these observations, also line-scans along the line perpendicular to the nodal line and along the nodal line were performed. The scans (10 line-scans averaged) perpendicular to the nodal line (horizontal dashed white line in 6(b); results shown in Fig.6(c)) reveals a weaker response of the particle when sitting in one of the main lobes (see inset in 6(c)). In this case, the particle is interacting with a homogeneous and linear electric field which is parallel to the SRR arms. For this configuration the SRR does not feature a resonance as already demonstrated in Fig. 3(b). If the nano-structure is moved toward the nodal line, the response measured in reflection (as well as transmission; not shown here) is still of considerable magnitude. In Fig. 6(d) the results for a scan along the nodal line are shown. The black curve represents the measured and averaged data (55 line-scans averaged) normalized to the reflection at the air-glass interface. The red curve shows the result of the iterative nonlinear least-squares fit routine described above. For the nodal line scan the response of the particle is stronger compared to the case before where the SRR arms were pointing perpendicular to the polarization of the incoming field. Additionally, the maximum of the response on the nodal line appears not centered (the maximal effect is not observed for the SRR sitting on the optical axis). Using an iterative nonlinear least-squares fit routine, we obtain a maximum amplitude of the resonant coupling effect of $B_{nl} = 0.000947(88)$.

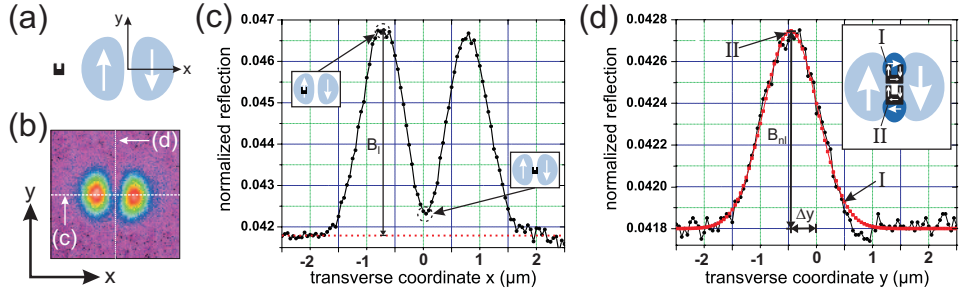

Fig. 6. Same as Fig. 5 but for a different orientation of the single SRR relative to the electric field of the incoming beam (SRR arms parallel to the electric field; see (a)). The inset in (d) shows a sketch of two coupling scenarios achieved for a scan along the nodal line. The additional overlap with the crossed electric field components on the nodal line (dark blue areas) leads to a shift ($\Delta y$) of the position of the observed maximum.

When the single SRR is moved along the nodal line the electric field of the main lobes overlapping with the SRR is not homogeneous across the nano-structure plane anymore. Coming closer to the optical axis the overlap with the z-component of the magnetic field and therefore with the curl of the electric field in focal plane increases. For the defined orientation of the SRR, the electric field can couple resonantly to the arms of the SRR with a phase shift of pi, thus driving eddy currents flowing around the nano-structure. When the scan is performed along the nodal line the maximum overlap with the magnetic field as well with the electric field in the lobes is obtained on the optical axis. Nevertheless, the observed maximum is shifted by $\Delta y = 450 nm \pm 100 nm$ in the negative y-direction (see Fig. 6(d)) and not observed at the optical axis (at around $y = 0 \mu$m). This small, but still significant shift can only be explained by an interaction of the SRR with the crossed electric field components being present on the nodal line above and below the optical axis (see Fig. 1(c)). The inset in Fig. 6(d) sketches the two situations of the SRR sitting on the nodal line several 100 nm above (I) and below (II) the optical axis in the focal plane. The dark blue areas represent the main electric field components in the focal plane. The light blue areas demonstrate the weak crossed electric field components with a maximum on the nodal line. The white arrows depict a snapshot of the electric field direction. For situation II the SRR already overlaps with the z-component of the magnetic field which is at maximum on the optical axis. Equivalently, the SRR arms overlap with the transverse electric field on the left and right hand side of the nodal line. Therefore the electric field can couple to the SRR arms with a phase shift of pi. Additionally, the crossed electric field component below the optical axis (negative y-direction) in the focal plane overlaps with the lower bar of the SRR, supporting the eddy current induced by the electric field of the main lobes overlapping with the SRR. By contrast, if the single SRR is positioned above the optical axis on the nodal line (situation I in Fig. 6(d)) the situation is changed. The crossed electric field component being present at this position also couples the lower bar of the SRR, but now it counteracts the current induced by the electric field of the main lobes overlapping with the SRR. Therefore, the effect in the scan data observed at this location is weaker compared to situation II. In summary, we attribute the observed shift of the maximal response to the additional coupling of the crossed electric field components by local enhancement or weakening of the induced eddy current for the given orientation of the SRR.

To proof this explanation qualitatively, we repeated these measurements with the same single SRR being rotated by 180 degrees in the x-y-plane (SRR arms pointing in the negative y-direction; results not shown here). In this case the maximal response appeared shifted in the positive y-direction relative to the optical axis. This is in line with the proposed explanation. In this coupling scenario, the upper crossed electric field component is supporting and the lower one is counteracting the induced eddy current. Therefore, the maximum is shifted to the other side relative to the optical axis.

If the SRR is placed between the lobes with its arms pointing into the y-direction the acting field is weak but has the perfect shape to excite the magnetic resonance. In contrast the strong field in the lobes points perpendicular to the bar of the SRR thus exciting the nano-structure with much lower efficiency. To compare the values of maximum scattering for the two situations, the overlap with the different electric energy densities present at the specific positions of the nano-structure relative to the beam have to be taken into account. By calculating the values of the overlapping electric energy density for both situations from the simulation data of the focal plane (see Fig. 1) and normalizing the measured maximal amplitudes for the excitation of the particle at the two positions in the beam, one yields a ratio of $B_{l,norm}/B_{nl,norm} = 0.132$. $B_{l,norm}$ and $B_{nl,norm}$ are defined as the maxima of the reflection signals observed for scans perpendicular to the nodal line and along the nodal line normalized to the corresponding fraction of the electric field energy incident on the metalized split ring area (see [29] for further details). This ratio is in good agreement with the ratio for non-resonant (electric field perpendicular to the SRR arms) and resonant (electric field perpendicular to the SRR arms) coupling at a wavelength of 1.525 $\mu$m derived from the resonance spectra recorded with a highly focused linearly polarized Gaussian beam (see Fig. 3(a) and (b); ratio: 0.233).

These experimental scan results state clearly, that the strength and type of interaction between the highly focused polarization mode and the single nano-structure strongly depend on the the position of the nano-structure and its orientation relative to the beam. In this context, the chosen polarization mode as well as the asymmetric shape of the SRR give access to different coupling scenarios with the electric field distribution in the focal plane. If the measured maximal responses to different field distributions are a consequence of the investigated so-called magnetic resonance around 1.525 $\mu$m, then the strength of the response should weaken if the wavelength is tuned away from this resonance. Note, that depending on the mode of excitation (resonant or non-resonant) the observed effects in reflection can weaken or increase, if the wavelength is changed. Following this approach we also performed the described scan measurements on the same SRR using a wavelength of 1.4 $\mu$m (the SRR is not resonant at this wavelength for both orientations; see Fig. 3). Again, both orientations of the SRR relative to the main electric field components in the focus were investigated. Note, that the diameter of the incoming beam at the entrance pupil of the MO was chosen such, that the diameter of the main electric field component in the focus stays almost constant (the diameter of the incoming beam was reduced compared to the measurements before). This leads to a change in strength of the field components of the electric and magnetic field in the focal plane. Furthermore, the normalization level (reflection at the air-glass interface) changes when the wavelength is varied in the setup. These effects are taken into account when comparing the results for different wavelengths.

Figure 7(b) - (d) show the scan results at a wavelength of 1.4 $\mu$m. Again, a y-polarized

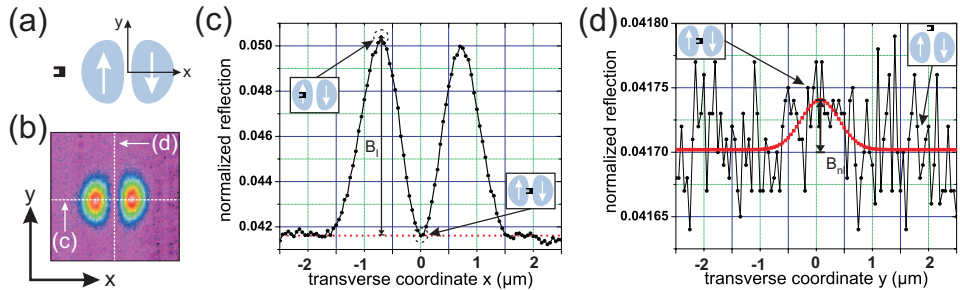

Fig. 7. Same as Fig. 5 but for a different wavelength of 1400 nm. The SRR is oriented as depicted in (a).

$TEM_{10}$-mode was used for investigations. The orientation of the single SRR was chosen as indicated in Fig. 7(a). In good agreement with the predictions derived from the resonance spectra (see Fig. 3), the maximal strength of the response decreases, when the SRR is positioned in the maximum electric field of one of the main lobes in the focus. The orientation of the SRR was chosen such, that the electric field points perpendicular to the SRR arms, therefore the coupling is not resonant anymore, because the wavelength was tuned away from the resonance resulting in a drop of the measured response (see Fig. 7(c)). The same holds for the response of the nano-structure, when it is placed on the nodal line between the lobes. To quantify this response, the nodal line was scanned 66 times. The measurements were averaged and normalized to the reflection at the air-glass interface. The results are shown in Fig. 7(d). Also for this coupling scenario the response weakens in comparison to the resonant case. Applying the iterative non-linear least-squares fit routine described above to the measurement data gives an amplitude of $B_{nl} = 0.000039(96)$ for the given situation.

If the SRR is rotated by -90 degrees, the possible coupling scenarios change again. The SRR

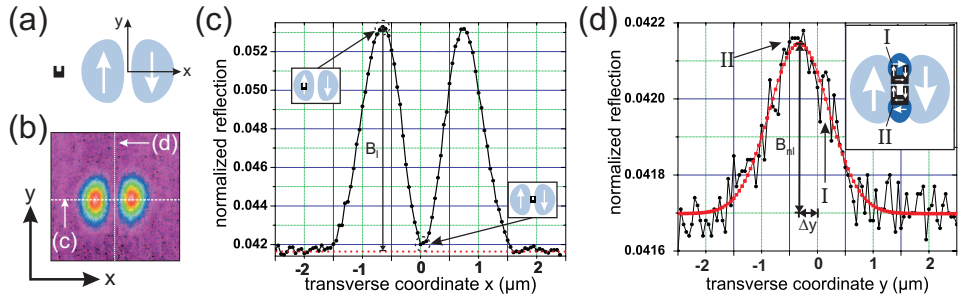

Fig. 8. Same as Fig. 7 but for a different orientation of the single SRR relative to the electric field of the incoming beam (SRR arms parallel to the electric field; see (a)). The wavelength was set to 1400 nm.

arms are now oriented parallel to the electric field of the incoming beam (see Fig. 7(a)). The response of the SRR sitting in one of the main lobes should therefore increase, due to the fact that the chosen wavelength is closer to the resonance observed around 1.25 $\mu$m (see Fig. 3(b)). However, the response of the SRR moved along the nodal line should weaken, because this coupling scenario is non-resonant at 1.4 $\mu$m. These tendencies are verified by our experiments. The results are shown in Fig. 7(b) - (d). The response of the SRR in one of the lobes is now stronger compared to the coupling strength achieved at a wavelength of 1.525 $\mu$m (see Fig. 7(c)). On the contrary, the response of the nano-structure when moved along the nodal line (overlapping with a magnetic field perpendicular to the SRR plane or equivalently an in-plane component of the electric field) is decreased (see Fig. 7(d)). The maximum effect for the SRR sitting on the nodal line for the given orientation is calculated by the nonlinear least-squares fit routine to be $B_{nl} = 0.00045(11)$. Moreover, the shift of the maximum is still observed but is smaller in comparison to the resonant case at 1.525 $\mu$m ($\Delta y = 230 nm \pm 100 nm$). By normalizing the measured maximal amplitudes for the excitation of the SRR at the two positions in the beam again with the values of the corresponding electric energy density (see [29]) at a wavelength of 1.4 $\mu$m one yields a ratio of 0.75 ($B_{l,norm}/B_{nl,norm}$), which is in good agreement with the resonance spectra as well (ratio from spectra: 0.81).

To compare the results for a wavelength of 1.4 $\mu$m to those at 1.525 $\mu$m also quantitatively, one has to take into account the respective field distributions in the focal plane and their overlap with the SRR for the different orientations and wavelengths. First, the response was measured for the two following geometries: 1) the SRR is sitting in one of the main lobes of the focused beam, the electric field being homogeneous across the SSR and oriented perpendicular to the SRR arms. 2) On the other hand the SSR is rotated by 90 degrees (SRR arms point upwards) and placed on the nodal line where the electric field is inhomogeneous coupling pi out of phase to the two arms. In both cases the results measured at the two wavelengths 1.4 $\mu$m and 1.525 $\mu$m scale just the same as expected from the spectrum of the 1.525 $\mu$m resonance in Fig. 3(a) (solid red stars and green hexagons in Fig. 9(a)). This shows, that in both geometries the 1.525 $\mu$m resonance dominates the coupling of the light to the SRR. In turn these geometries can be used to study this resonance. Second, the response was measured for a third geometry: 3) the SRR is sitting in one of the main lobes of the focused beam, the electric field being homogeneous across the SSR and oriented parallel to the SRR arms. Now the results measured at the two wavelengths 1.4 $\mu$m and 1.525 $\mu$m scale just the same as expected from the spectrum of the 1.4 $\mu$m resonance in Fig. 3(b) (solid red stars in Fig. 9(b)). This shows that in geometry 3) the 1.4 $\mu$m resonance determines the coupling. In geometry 2) an eddy current is induced in the SRR by the two electric field components oscillating pi out of phase, i.e. by the curl of the electric field on the nodal line or equivalently the magnetic field on the nodal line normal to the SRR plane. As a result, the observation that the normalized reflections in Fig. 6 and 8 does not go back to the background level between the lobes is clearly an effect of the magnetic dipole contribution to the 1.525 $\mu$m resonance.

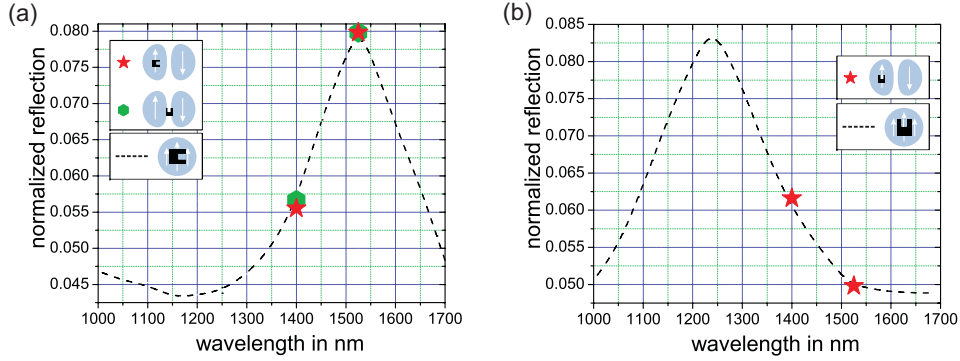

Fig. 9. Comparison between the resonance spectra (dashed black line; reflection only) of the tested SRR and the scan results achieved for coupling with different field distributions present in the focus of a y-polarized $TEM_{10}$-mode (solid red stars: SRR sitting in one of the main lobes; solid green hexagon: SRR sitting on the nodal line with arms pointing upwards). The presented data points resulting from the scan measurements are renormalized to the overlap with the electric field energy density in the focus for different wavelengths. For each of the three geometries shown in the insets the reflection was measured at two different wavelengths (green and red data points). For each geometry the data point measured at 1.525 $\mu$m and normalized to the incident electric field energy was scaled to fit to the resonance spectrum (dashed line). The data point at 1.4 $\mu$m was plotted using the same scale factor. The resonance spectra were recorded with a linearly polarized Gaussian beam. (a) Spectra taken for linear electric field aligned perpendicular to the SRR arms. (b) For the electric field parallel to the SRR arms.

If we would have an idealized split ring with a circular shape and only a narrow gap, the coupling to the normal magnetic field component would not depend on the orientation of the gap. The observation that the normalized reflection in Fig. 5 and 7 comes close to the background level between the lobes is a result of the non-ideal shape of the split ring and the electromagnetic field not being rotationally symmetric. Still, it does not quite reach the background level exactly and this remaining effect, although barely significant (see $B_{nl}$ in Fig. 5(d) and 7(d) scales with wavelength like the 1.525 $\mu$m resonance line in Fig. 3(a).

## 6. Conclusion

In conclusion, we have presented a versatile experimental setup and measurement technique which can be used to investigate the optical properties of single sub-wavelength nano-structures. We have demonstrated our technique using a single split-ring resonator as a model system. After having identified the spectral positions of the resonances of the nano-structure, we investigated the so-called magnetic resonance in detail. By measuring the spatial distribution of the field pattern re-radiated by the SRR, we have verified the non-vanishing magnetic dipole contribution of the so-called magnetic resonance. The internal dynamics of the resonance was studied using highly focused polarization tailored light beams with the full freedom in choosing the pattern of the exciting electric vector field.

The new experimental technique presented here can help to identify the response of other nano-structures especially if several resonances exist resulting in a more complex response. By adapting the polarization pattern of the excitation field to the nano-structure or its resonances under test, the presented measurement techniques can give a detailed insight into the physics of the resonances of a single sub-wavelength nano-structure.

**Acknowledgment**

We acknowledge the help of Daniel Ploß, who fabricated the investigated samples. Furthermore, we thank Norbert Lindlein for his help with the simulations. We thank Martin Wegener, Stefan Linden and Nils Feth for stimulating discussion. Additionally, we acknowledge the support by the Cluster of Excellence for Engineering of Advanced Materials (EAM) located in Erlangen, Germany.